\documentstyle[11pt,newpasp,twoside,epsf]{article} 
\def\edcomment#1{\iffalse\marginpar{\raggedright\sl#1\/}\else\relax\fi} 
\reversemarginpar

\def\plotfour#1#2#3#4{\centering \leavevmode 
\epsfxsize=.35\columnwidth \epsfbox{#1} \hfil 
\epsfxsize=.35\columnwidth \epsfbox{#2} \hfil 
\epsfxsize=.35\columnwidth \epsfbox{#3} \hfil 
\epsfxsize=.35\columnwidth \epsfbox{#4}} 
 
\def\plotsix#1#2#3#4#5#6{\centering \leavevmode 
\epsfxsize=.45\columnwidth \epsfbox{#1} \hfil 
\epsfxsize=.45\columnwidth \epsfbox{#2} \hfil 
\epsfxsize=.45\columnwidth \epsfbox{#3} \hfil 
\epsfxsize=.45\columnwidth \epsfbox{#4} \hfil 
\epsfxsize=.45\columnwidth \epsfbox{#5} \hfil 
\epsfxsize=.45\columnwidth \epsfbox{#6} }

\def\plotonea#1{\centering
\leavevmode \epsfxsize=.45\columnwidth \epsfbox{#1}}

\def\etal{{\it et al.\ }}

\def\eg{{e.g.,}} 
\def\ie{{i.e.,}}

\def\puncspace{\ifmmode\,\else{\ifcat.\C{\if.\C\else\if,\C\else\if?\C\else%
\if:\C\else\if;\C\else\if-\C\else\if)\C\else\if/\C\else\if]\C\else\if'\C%
\else\space\fi\fi\fi\fi\fi\fi\fi\fi\fi\fi}%
\else\if\empty\C\else\if\space\C\else\space\fi\fi\fi}\fi} 
\def\SP{\let\\=\empty\futurelet\C\puncspace }

\def\vnabla{\pmb{$\nabla$}} 
\def\div{\vnabla\!\cdot\!}

\def\divv{\div\vv}

\def\dnsig{$D_n-\sigma$\SP} 
\def\ltsima{$\; \buildrel < \over \sim \;$} 
\def\lsim{\lower.5ex\hbox{\ltsima}} 
\def\gtsima{$\; \buildrel > \over \sim \;$} 
\def\gsim{\lower.5ex\hbox{\gtsima}} 
\def\ga{\mathrel{\hbox{\rlap{\hbox{\lower4pt\hbox{$\sim$}}}\hbox{$>$}}}} 
\def\la{\mathrel{\hbox{\rlap{\hbox{\lower4pt\hbox{$\sim$}}}\hbox{$<$}}}} 
 
\def\ifm#1{\relax\ifmmode#1\else$\mathsurround=0pt #1$\fi} 
\def\kms{\,{\rm km\,s\ifm{^{-1}}}}

\def\la{\langle}

\def\pmb#1{\setbox0=\hbox{#1}%
 \kern-.025em\copy0\kern-\wd0 
 \kern.05em\copy0\kern-\wd0 
 \kern-.025em\raise.0433em\box0}

\def \vr {{ \bf r}} 
\def \br {{ \bf r}} 
\def \bv {{ \bf v}}

  \def\3hmpc{\, ( h^{-1} {\rm Mpc})^3} 
\def\kms{\, {\rm km\,s^{-1}}}   
\def\h5{h_{50}} 

\def \vv {{ \bf v}}

\def \WF {^{\rm WF}}

\def \r0p{ r{_0^\prime}}

\def \m3{{\rm Mark III}} 
 
\def\uo{u^o}

\begin{document}
\title{Reconstructing the ZOA from Galaxy Peculiar Velocities}
\author{Saleem Zaroubi} 
\affil{Max-Planck-Institu f\"ur Astrophysik,
Karl-Schwarzschild Str. 1, D-85741 Garching, Germany}
 
\begin{abstract}
Galaxy peculiar velocity data provide important dynamical clues to the
structures obscured by the Zone of Avoidance (hereafter, ZOA) with
resolution $\gsim 500\kms$. This indirect probe complements the very
challenging approach of directly mapping of the distribution of
galaxies behind the Milky Way.  In this work, the Wiener filter method is
applied to reconstruct the 3D density and velocity distributions of
the universe, within $\lsim 8000\kms$, using SEcat the largest
peculiar velocity catalog yet. This catalog is a combination of the
SFI spiral galaxy peculiar velocity catalog and the newly completed
nearby early-type galaxy peculiar velocity catalog, ENEAR. The
recovered density is smoothed with $900 \kms$ Gaussian. The main
reconstructed structures are consistent with those extracted from the
IRAS 1.2-Jy redshift galaxy catalogs. The revealed structures within the ZOA
are identified and their robustness and significance are discussed.
\end{abstract} 
 
\section{Introduction} 
\label{sec:intro}
Extinction due to the galactic plane obscures about $25\%$ of the
optically visible universe. In order to account for the Local Group
motion relative to the Cosmic Microwave Background (CMB), the flow of
galaxies in the Great-Attractor (hereafter, GA) area and other similar
phenomena, the full distribution of matter, especially in the local
universe, is essential. Direct measurement of the distribution of
matter/galaxies requires extensive, tedious and dedicated observational
programs in all the available electromagnetic wave-bands (see review
by Kraan-Korteweg \& Lahav 2000 and references therein).
 
However, a complementary approach for studying the universe behind the
Milky Way is to use the available dynamical data, \eg\, galaxy
peculiar velocity catalogs, together with statistical reconstruction
methods, \eg\, Wiener filtering (WF), in order to uncover the mass
density distribution, with resolution scale $\gsim 500\kms$, hence,
singling out the dynamically most significant structures.
 
Peculiar velocities of galaxies enable a direct and reliable
measurement of the {\it underlying} mass distribution, under the
natural assumption that galaxies are unbiased tracers of the
large-scale, gravitationally induced, velocity field. Furthermore,
since peculiar velocities are non-local and have contributions from
different scales and different regions, analysis of the peculiar
velocity field provides information on regions not covered by the data,
\eg\, the ZOA (Kolatt \etal 1995; Zaroubi \etal 1999), and on scales
larger than the sampled regions (Hoffman \etal 2000).
 
Kolatt \etal (1995) were the first to attempt to reconstruct the ZOA
from galaxy peculiar velocity data, where they used the POTENT method
(Bertschinger \& Dekel 1989; Dekel 1994) to reconstruct the
mass-density distribution, within a sphere of radius $8000\kms$, from
the \m3 galaxy peculiar velocity catalog (Willick \etal 1997) with
$1200\kms$ resolution. Their study has resulted in predicting that the
mass distribution of the Great Attractor peaks precisely at the center
of the ZOA at a distance of $\approx 4500\kms$.  Zaroubi \etal (1999)
have used the same peculiar velocity catalog to Wiener reconstruct the
mass density distribution within $8000\kms$ sphere.  Their main
conclusions were consistent with those of Kolatt \etal (1995).

The common use of WF is for straightforward noise suppression, but it
can be easily generalized to achieve two further goals: to reconstruct
the density field from the observed radial velocities and to
interpolate or extrapolate the reconstruction to regions of poor
sampling (for review, see Zaroubi \etal 1995 \& Hoffman 2000). The
later aspect is of special use for the reconstruction of the ZOA. In
this work the WF is used to reconstruct the mass density distribution
within $8000\kms$ sphere from the largest galaxy peculiar velocity
catalog yet. This catalog is a combination of the SFI and the ENEAR
peculiar galaxy catalog.
 
The WF approach has been already applied to the IRAS two and
three-dimensional galaxy distribution (Lahav \etal\ 1994), the IRAS
three-dimensional redshift distortions (Fisher \etal\ 1995; Webster
\etal 1997), the COBE/DMR cosmic microwave background mapping (Bunn
\etal\ 1994), and to galaxy peculiar velocity catalogs of \m3 and
ENEAR (Zaroubi \etal 1999; 2000a).


The outline of this paper is as follows. In \S~2 we briefly describe
the peculiar velocity data used in the present analysis. The method of
Wiener reconstruction from peculiar velocity data is introduced in 
\S~3, and the results of its application to the SEcat data set are
presented in \S~4. The paper concludes with a general discussion (\S~5).

\section{The Data Sets}
\label{sec:data} 

The ENEAR catalog have been extracted from the all-sky ENEAR redshift
survey comprising about 1600 galaxies.  Individual galaxy distances
were estimated from a direct \dnsig template relation derived by
combining all the available cluster data, corrected for incompleteness
and associated diameter-bias. From the observed scatter of the
template relation the estimated fractional error in the inferred
distance of a galaxy is $\Delta \sim 0.19$, nearly independent of the
velocity dispersion. An objective grouping procedure has been applied
to the data in order to lower the inhomogeneous Malmquist bias before
correction and to avoid strong non-linear effects (in particular large
velocities of galaxies in clusters). The final catalog consists of
about $750$ objects.

The SFI catalog of peculiar velocities of galaxies (Giovanelli \etal
1999), contains about 1300 field spiral galaxies with Tully-Fisher
distances. After the grouping procedure the final dataset consists of
distances, radial peculiar velocities and errors for $\approx 1250$
objects, ranging from individual field galaxies to rich clusters.

The combined catalog, SEcat, consists of $\approx 2000$ objects,
uniformly covering, apart from the ZOA region, the local universe up to
distance of $\approx 6000\kms$. The error in the distance of the
objects measured with \dnsig, namely objects from the ENEAR catalog,
are assumed to have two contributions, the first is the usual \dnsig
distance proportional errors. The second is a constant error of $250
\kms$ that accounts for the non-linear velocities of galaxies in the
high density environment in which early-type galaxies preferentially
reside.

The inferred distances are corrected for the homogeneous and
inhomogeneous Malmquist bias (for details see Freudling \etal 1999; da Costa
\etal\ 2000b). The latter was estimated using the PSCz
density field (Branchini \etal 1999), corrected for the effects of
peculiar velocities, using the expressions given by Willick \etal
(1997). In this calculation, a correction for redshift limit of the
sample is included.
 
Finally the results are compared with the mass-density reconstruction
from the \m3 catalog (Willick \etal\ 1997). This catalog, consists of
more than 3400 galaxies, has been compiled from several data sets of
spirals and elliptical/S0 galaxies with distances inferred by the
forward Tully-Fisher and $D_n-\sigma$ distance indicators.  These data
were re-calibrated and self-consistently put together as a homogeneous
catalog for velocity analysis.  The catalog provides radial velocities
and inferred distances with errors on the order of $17-21\%$ of the
distance per galaxy. After grouping, the catalog contains $\approx
1200$ objects. The sampling covers the whole sky outside the ZOA, but
with an anisotropic and non-uniform density that is a strong function
of distance. The good sampling typically ranges out to $6000\kms$ but
it may be limited to only $4000\kms$ in some directions or extend
beyond $8000\kms$ in other directions. The inhomogeniety of the \m3
sampling together with the complicated calibration proceedure employed
to obtain the final catalog led many authors to question its
reliability (\eg\, Davis \etal 1996).

\section{Wiener Filter} 
\label{sec:wiener} 
 
Here we limit the description of the WF to the actual application of
the method to the case of radial velocity data.  The data for the WF
analysis are given as a set of observed radial peculiar velocities
$\uo_i$ sampled at positions $\vr_i$ with estimated errors
$\epsilon_i$ that are assumed to be uncorrelated.  The peculiar
velocities are assumed to be corrected for systematic errors such as
Malmquist bias.  The observed velocities are thus related to the true
underlying velocity field $\bv(\br)$, or its radial component $u_i$ at
$\vr_i$, via
\begin{equation} 
\uo_i = \bv(\br_i) \cdot \hat \br_i  + \epsilon_i 
\equiv u_i + \epsilon_i . 
\label{eq:eps} 
\end{equation} 
 
We assume that the peculiar velocity field $\vv(\vr)$ 
and the density fluctuation 
field $\delta(\vr)$ are related via linear gravitational-instability theory, 
$\delta = f(\Omega)^{-1} \divv$, where $f(\Omega)\approx\Omega^{0.6}$ 
and $\Omega$ is the mean universal density parameter. 
Under the assumption of a specific theoretical prior for the power 
spectrum $P(k)$ of the underlying density field, 
we can write the WF minimum-variance estimator of the fields as 
\begin{equation} 
\bv\WF(\br) = \Bigl < \bv(\br) \uo_i \Bigr >  \Bigl < \uo_i \uo_j \Bigr > 
^{-1}        \uo_j 
\label{eq:WFv} 
\end{equation} 
and 
\begin{equation} 
\delta\WF(\br) = \Bigl < \delta(\br) \uo_i \Bigr >  \Bigl < \uo_i \uo_j 
\Bigr > ^{-1}    \uo_j  . 
\label{eq:WFd}
\end{equation} 
In these equations $ \Bigl < ... \Bigr > $ denotes an ensemble average.
The assumption that linear theory is valid on all scales enables us to
estimate, given the power spectrum, the ensemble average quantities
appearing in Eqs. (2) \& (3) . The reader is referred
to Zaroubi \etal (1999) for the explicit mathematical formulae used in
the calculation. We choose to reconstruct the density field with a
finite Gaussian smoothing of radius $900\kms$.

\section{Results}
\begin{figure} 
\plotonea{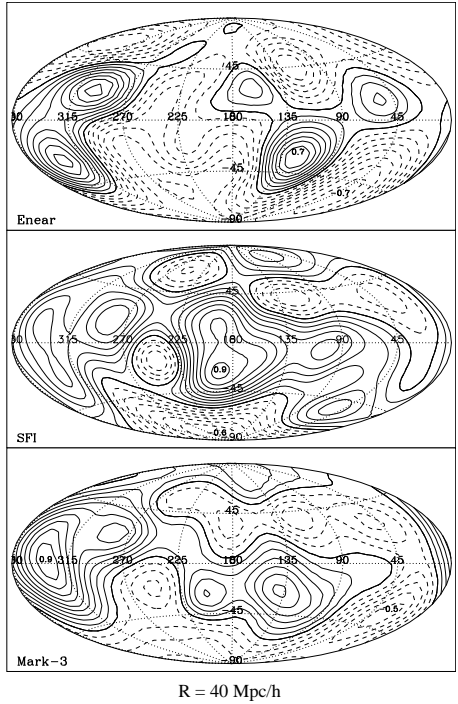} 
\caption{The reconstructed WF density fields from ENEAR (upper panel),
SFI (middle panel) and MARK III (lower panel) catalogs shown on
spherical shell at $4000\kms$ distance. The Aitoff projection is shown
in Galactic coordinates ($l,b$).  Density contour spacing is 0.1,
positive contours are solid, negative contours are dashed and
$\delta=0$ is denoted by heavy- sold line}
\label{fig:compare40} 
\end{figure} 
 First we compare the density reconstruction from the \m3, SFI and
ENEAR catalogs. Figure 1 shows the reconstructed mass-density
distribution for each catalog, smoothed with a $900\kms$ Gaussian, on a
spherical shell at $4000\kms$ distance. The assumed power spectrum
used in the reconstruction from \m3, SFI \& ENEAR has been
determined through maximum likelihood analysis by Zaroubi \etal
(1997), Freudling \etal (1999) and Zaroubi \etal (2000a), respectively.

In the three maps the existence of the GA supercluster on the left and
the Perseus-Pisces (P-P) supercluster to the right,
$(l,b)\approx(135\deg,-30\deg)$ is evident. However, the ENEAR map,
relative to the two others, shows a more localized GA and P-P. In
fact, since the ENEAR catalog measures velocities of early-type
galaxies preferentially residing in high density environments, this
difference is expected. Conversely, smaller overdensities for the GA
and P-P shown in the SFI density reconstruction map is due to the
tendency of spiral galaxies to reside in the field. The insufficient
sampling of the P-P supercluster in the \m3 catalog renders its
recovered density smaller than that of the GA.
\begin{figure}
\plotfour{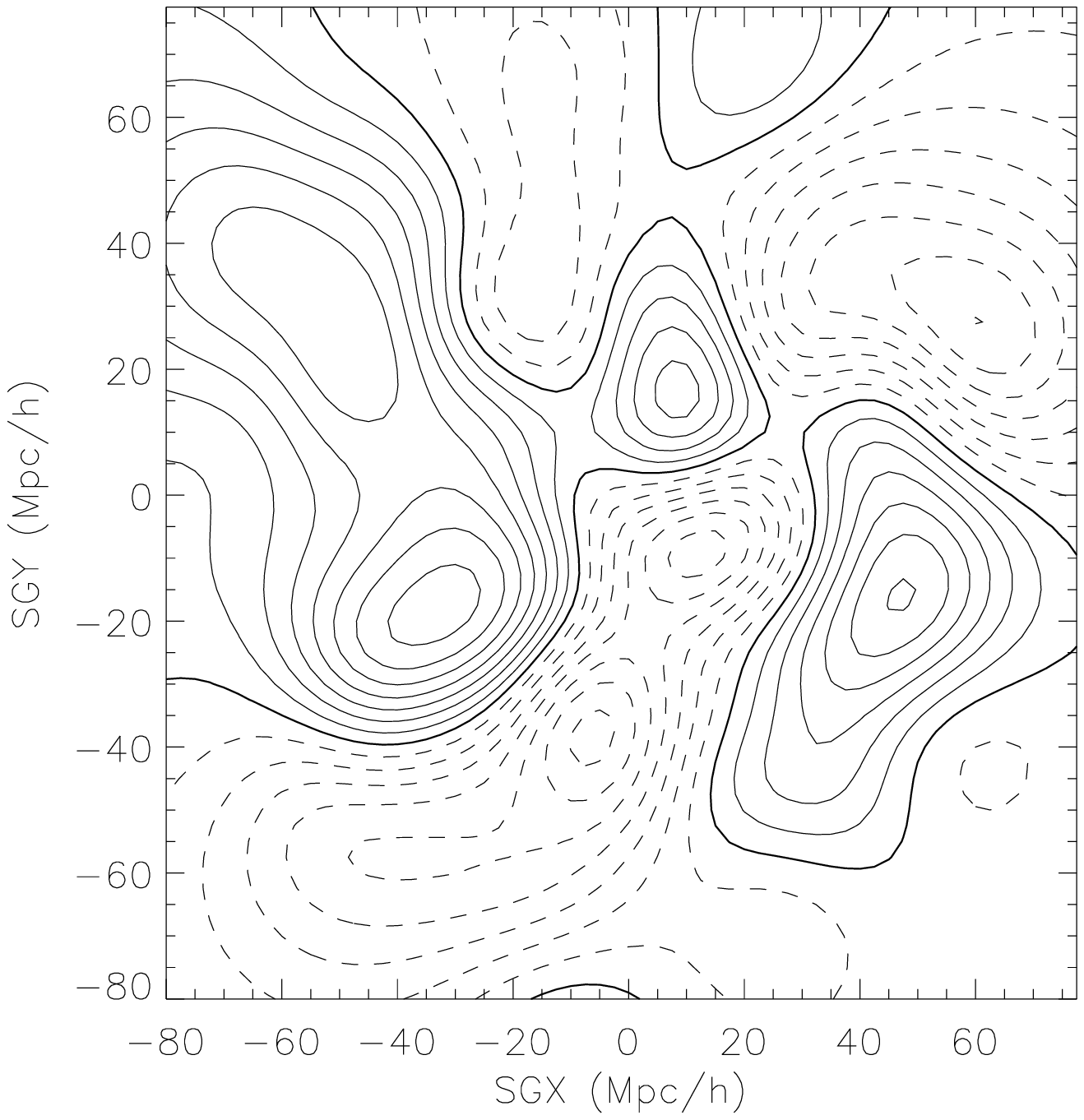}{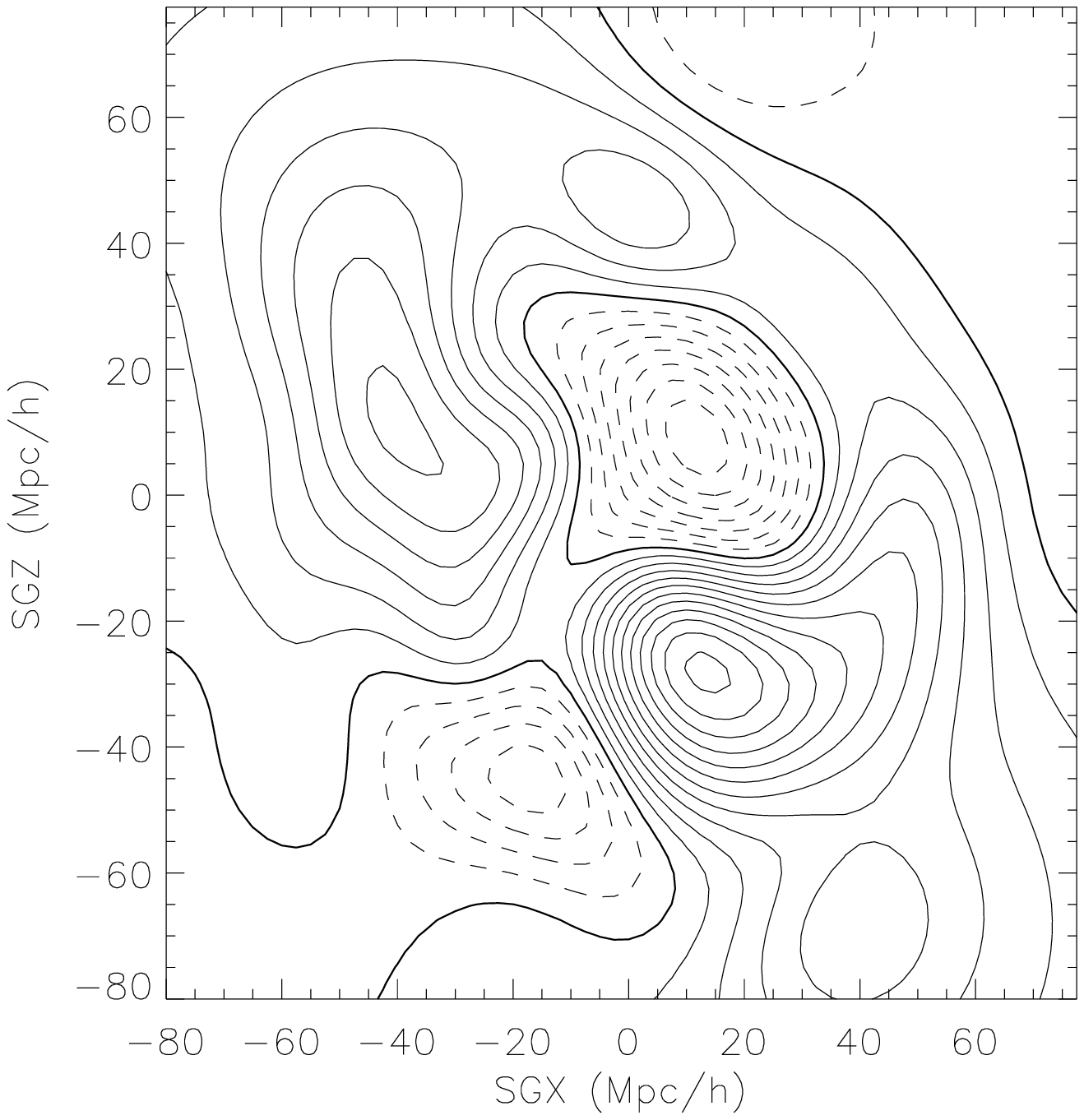}{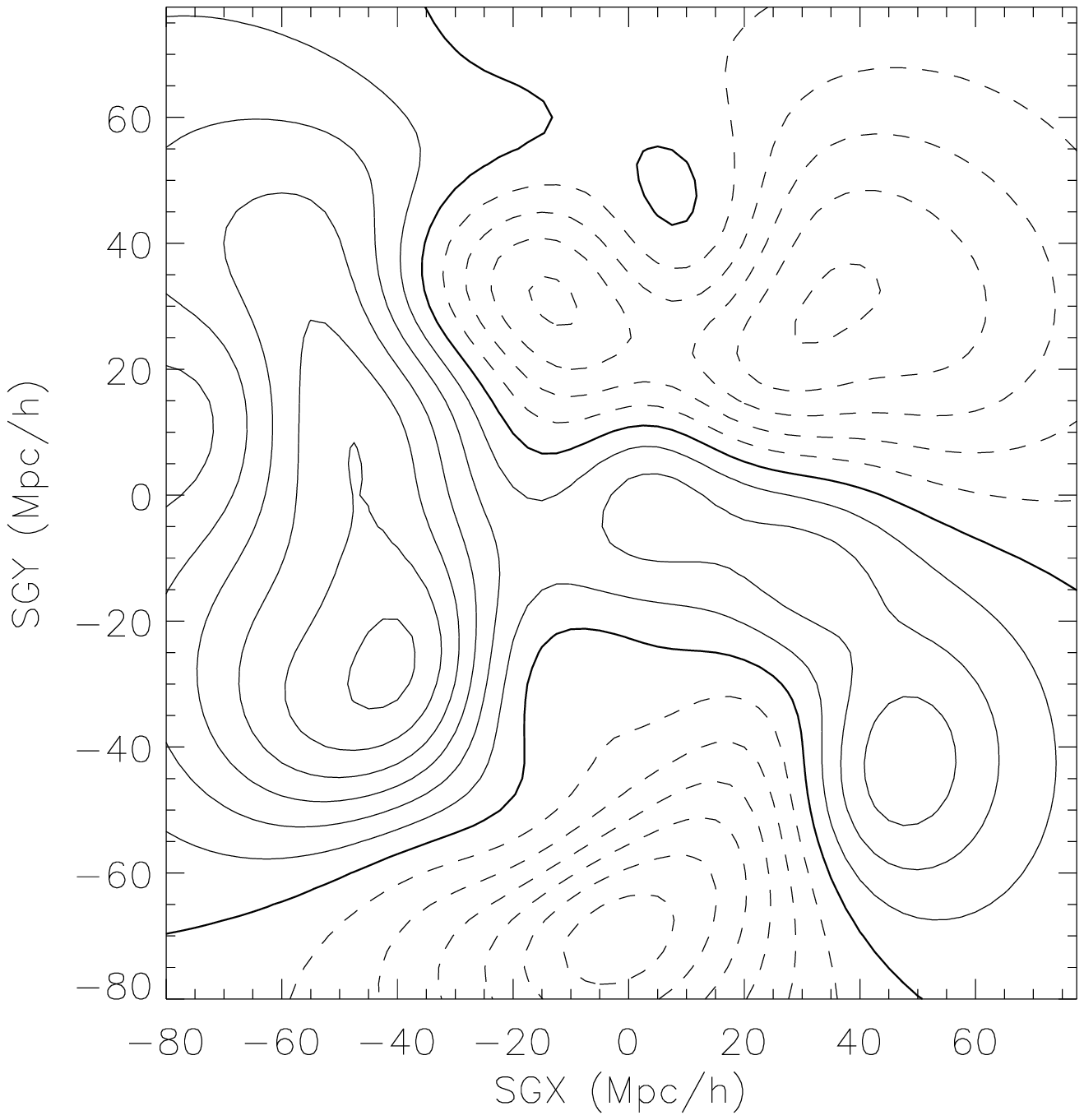}{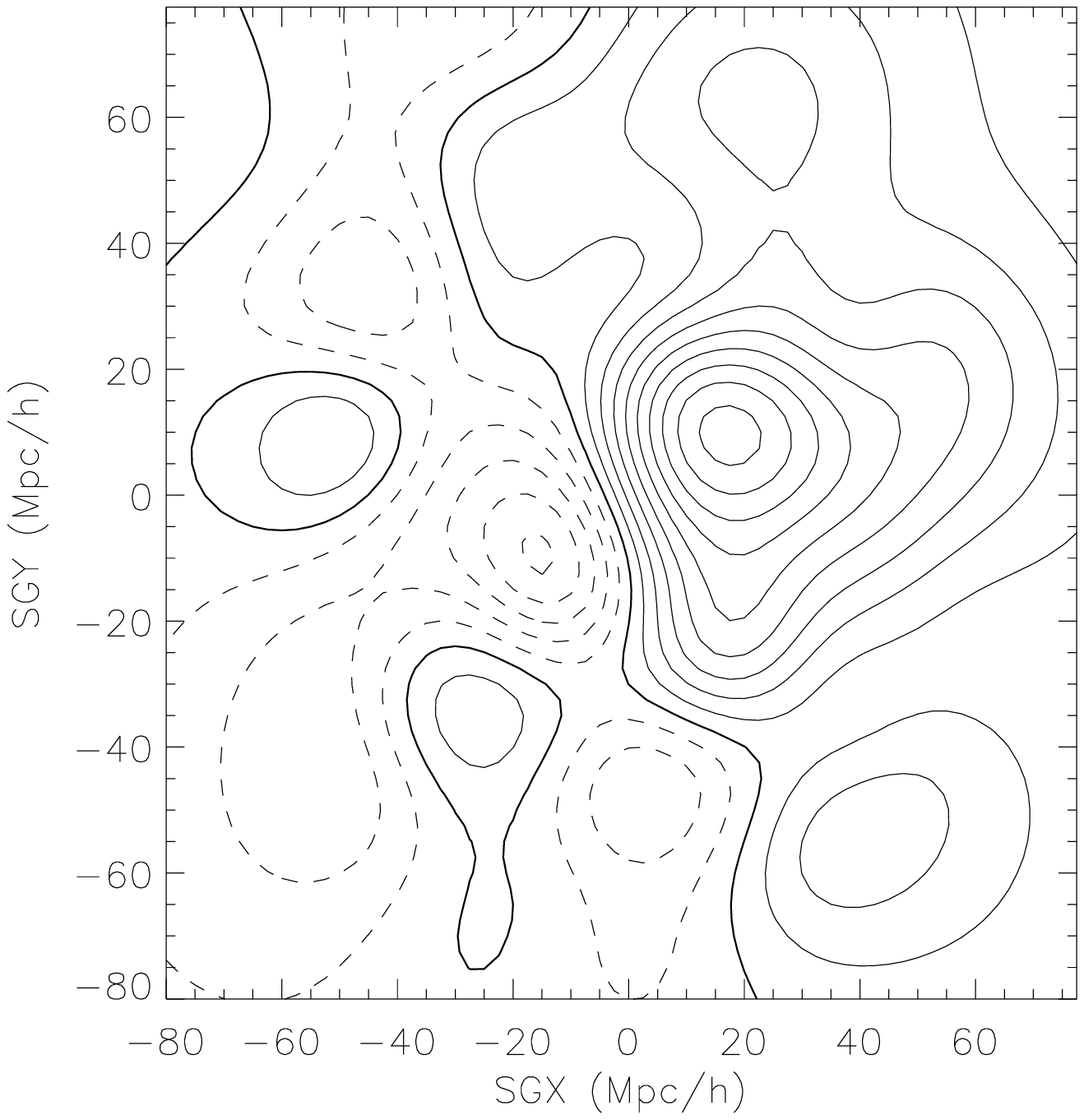}
\caption{The WF reconstructed density maps from the SEcat catalog. Top
left: The density field map in the Supergalactic (SGZ=0) plane, with
$900\kms$ Gaussian smoothing.  Density contour spacing is 0.1,
positive contours are solid, negative contours are dashed and
$\delta=0$ is denoted by heavy-solid line. Top right: the same as the
previous panel but for the $SGY=0$ plane , this plane mostly
coincides with the ZOA. Bottom left: The same as before but for the
SGZ$=4000\kms$ plane. Bottom right: The same as before but for the
SGZ$=-4000\kms$ plane.}
\label{fig:SG}
\end{figure}

The big structure, centered at $(l,b)\approx (200\deg,-30\deg)$,
appearing in the SFI density reconstruction and in the correspondent
SEcat spherical shell (see Figure 3), does not have a counterpart in
the ENEAR density map and has much lower density in the \m3
reconstruction. Comparison with the IRAS 1.2-Jy redshift survey
density reconstruction (Webster \etal 1997) shows that the peak
location of this structure coincides with the position of the massive
cluster N1600. The IRAS 1.2-Jy $4000\kms$ shell further shows the
existence of several other clusters, \ie\, Cancer, Camelopardalis,
${\bf C}_\beta$, ${\bf C}_\gamma$, ${\bf C}_\delta $ and P-P that can
account for this huge concentration seen extended from $l\approx
180\deg-220\deg$ and centered around the ZOA.

Obviously, the ENEAR and SFI catalogs complement, therefore we
combined them to one catalog, SEcat.  Figure 2 shows the maps of the
density field, recovered rfom the SEcat catalog, in 4 different slices
using a Gaussian smoothing of $900\kms$. In the Supergalactic plane
slice (upper left) the main features of our local universe can be
easily identified, including the GA and the P-P superclusters at the
left and right parts of the map respectively; the Local supercluster
appears at the center of the map. The SGY=0 slice (upper right)
coincides roughly with the plane obscured by the ZOA. Another two
slices are at SGZ$=\pm 4000\kms$ are also shown.

\begin{figure} 
\plotsix{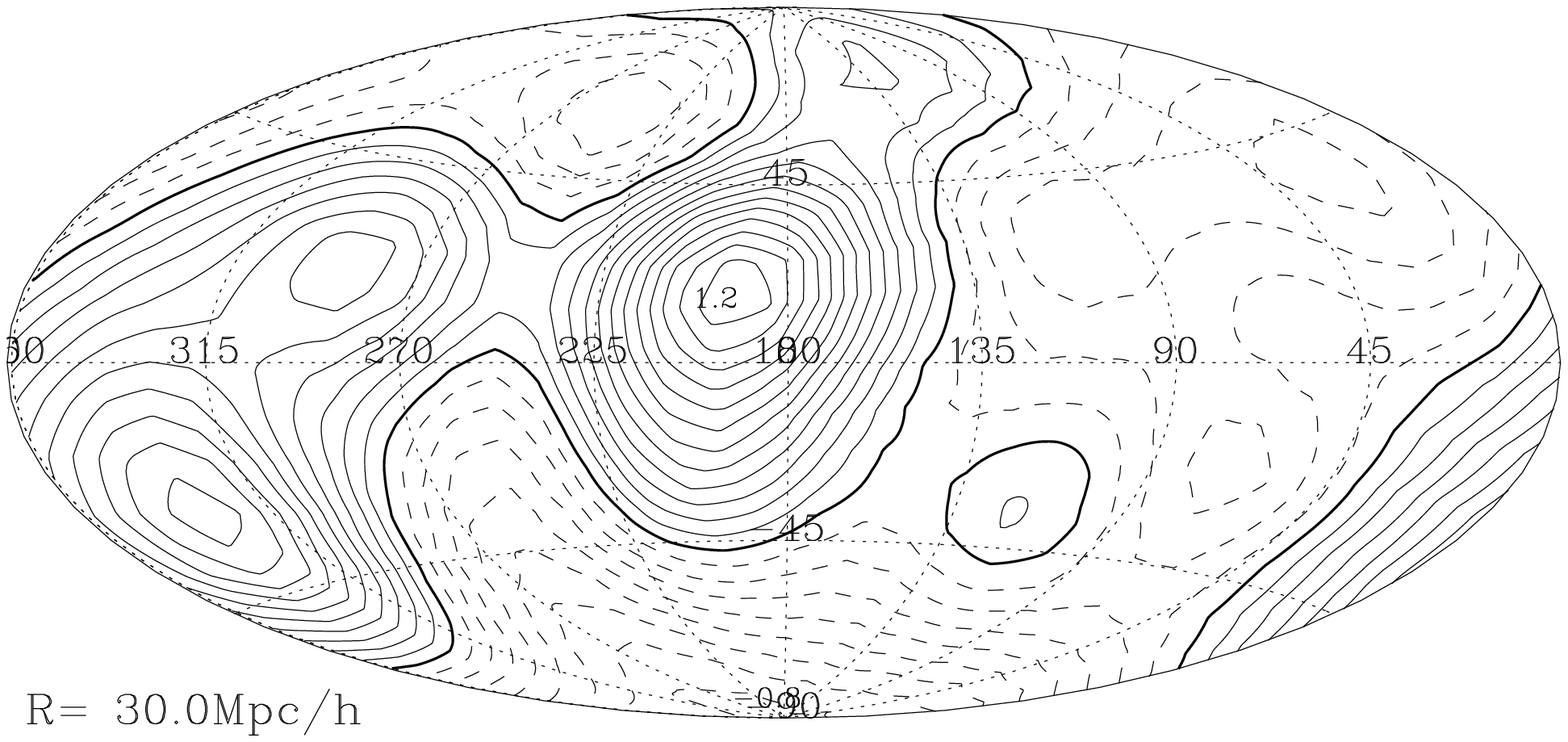}{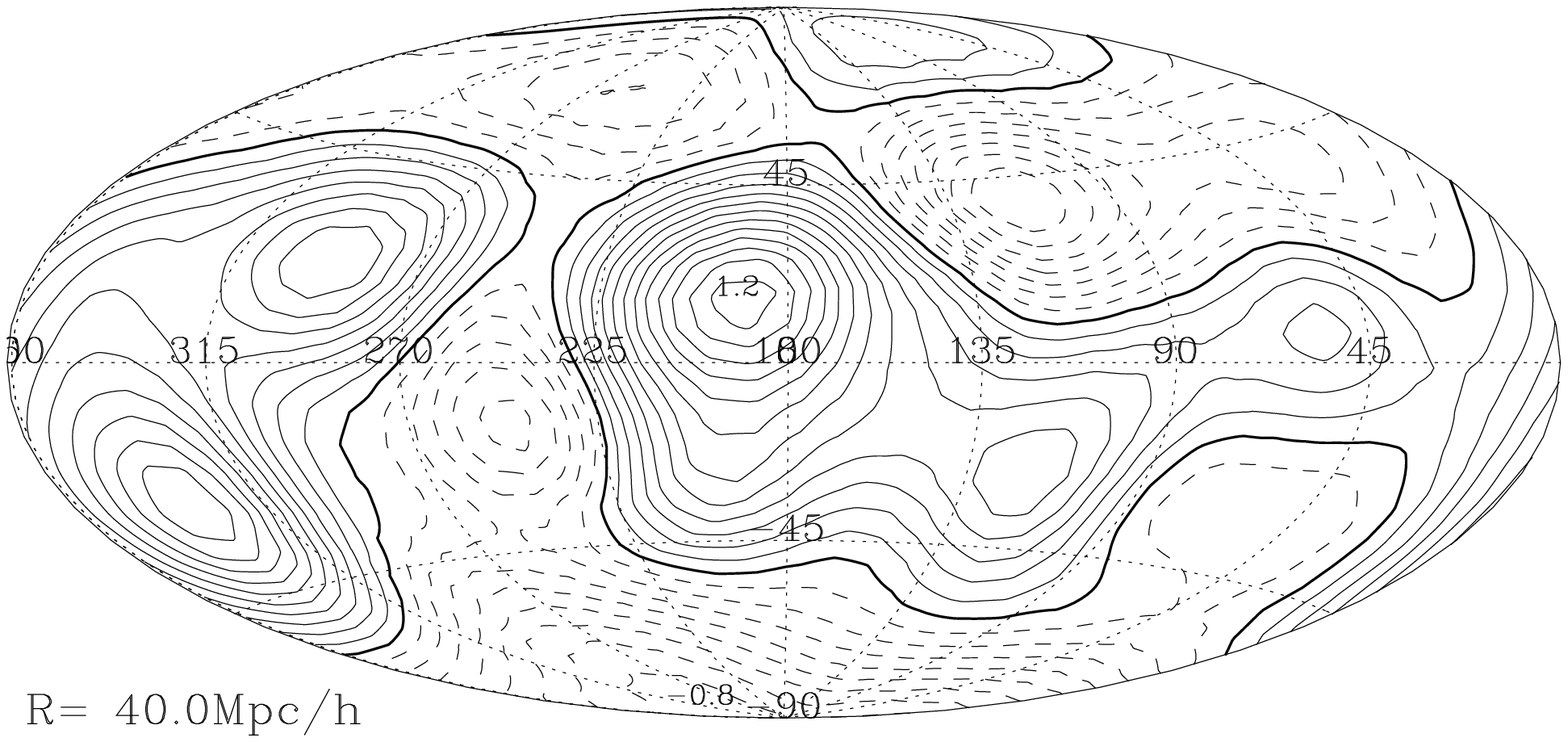}{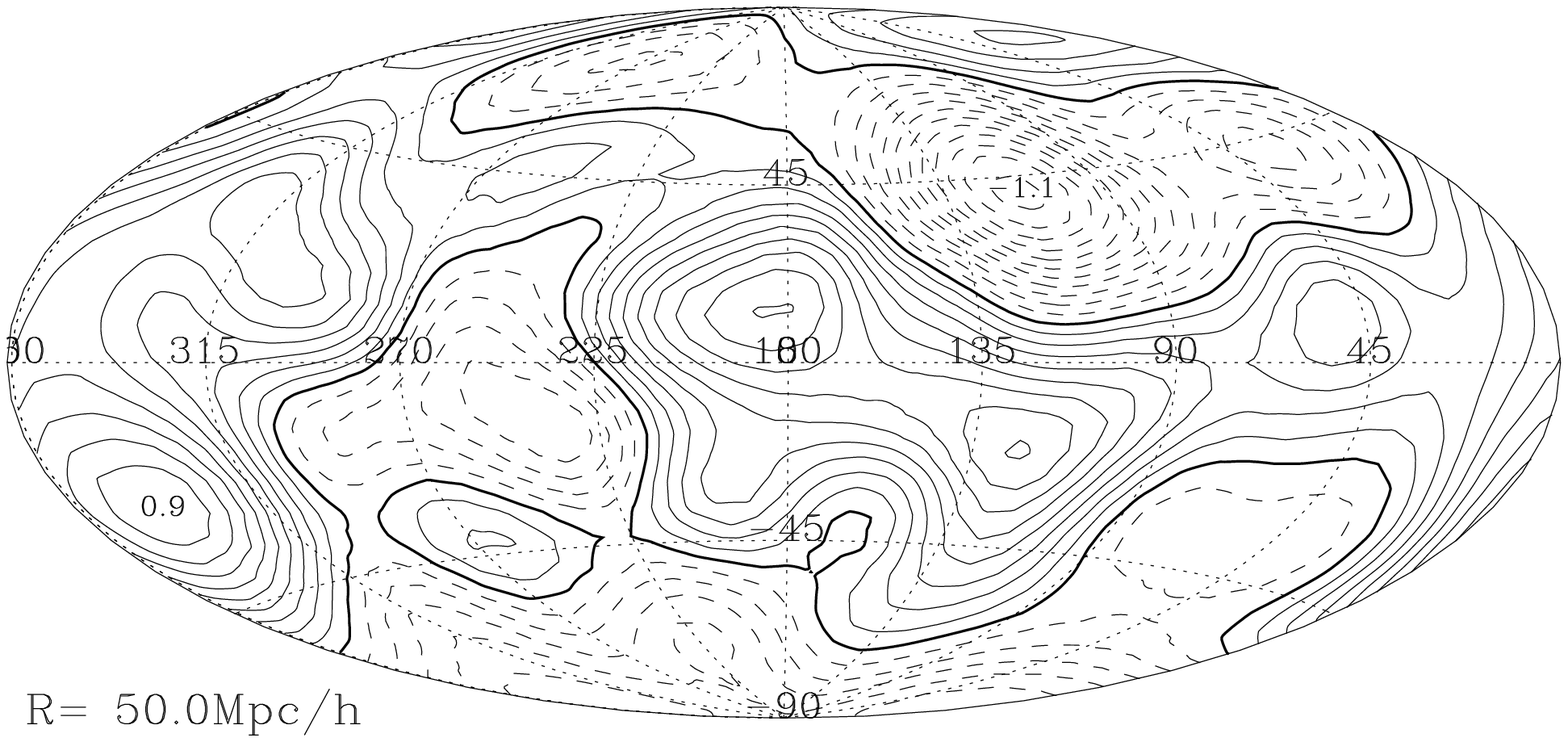}
        {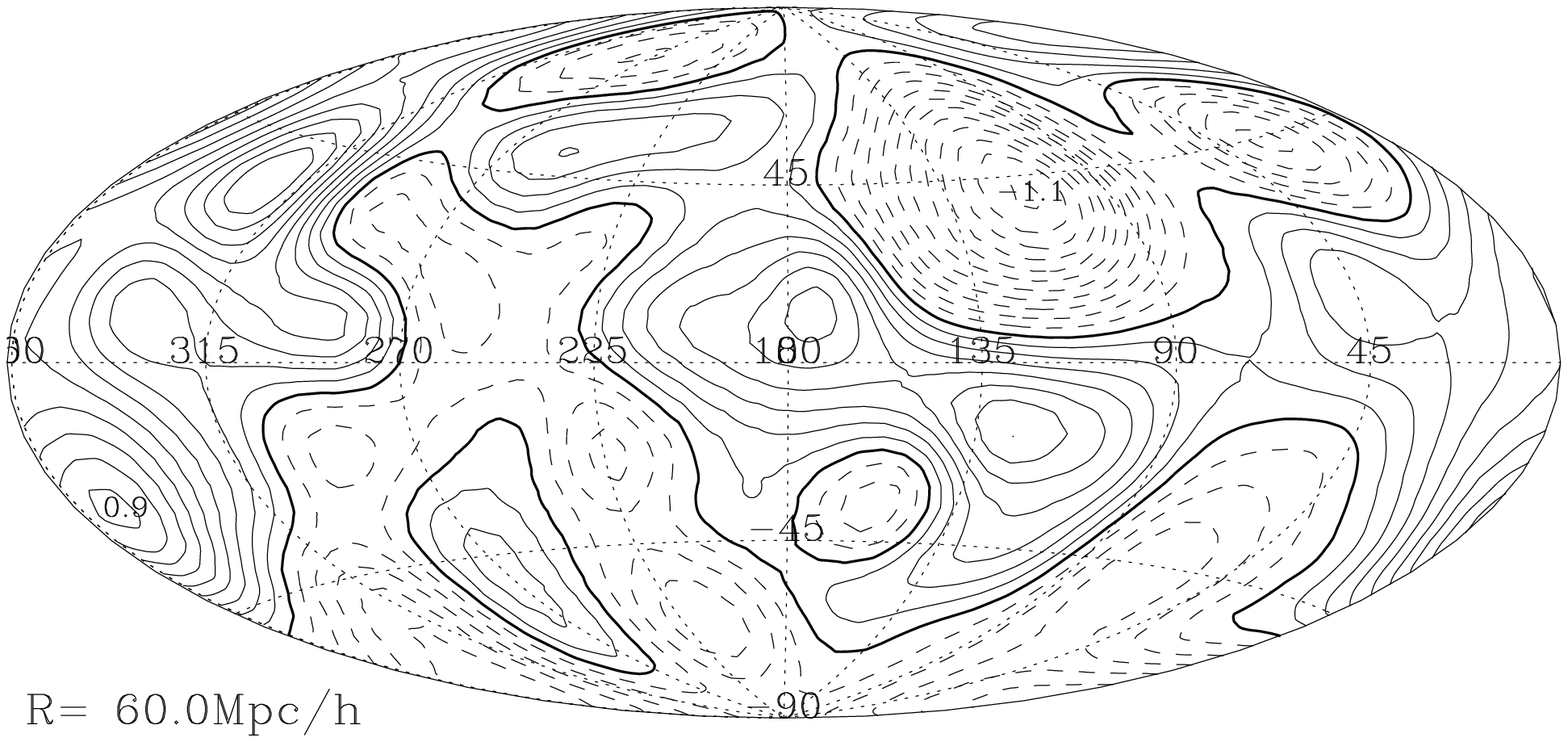}{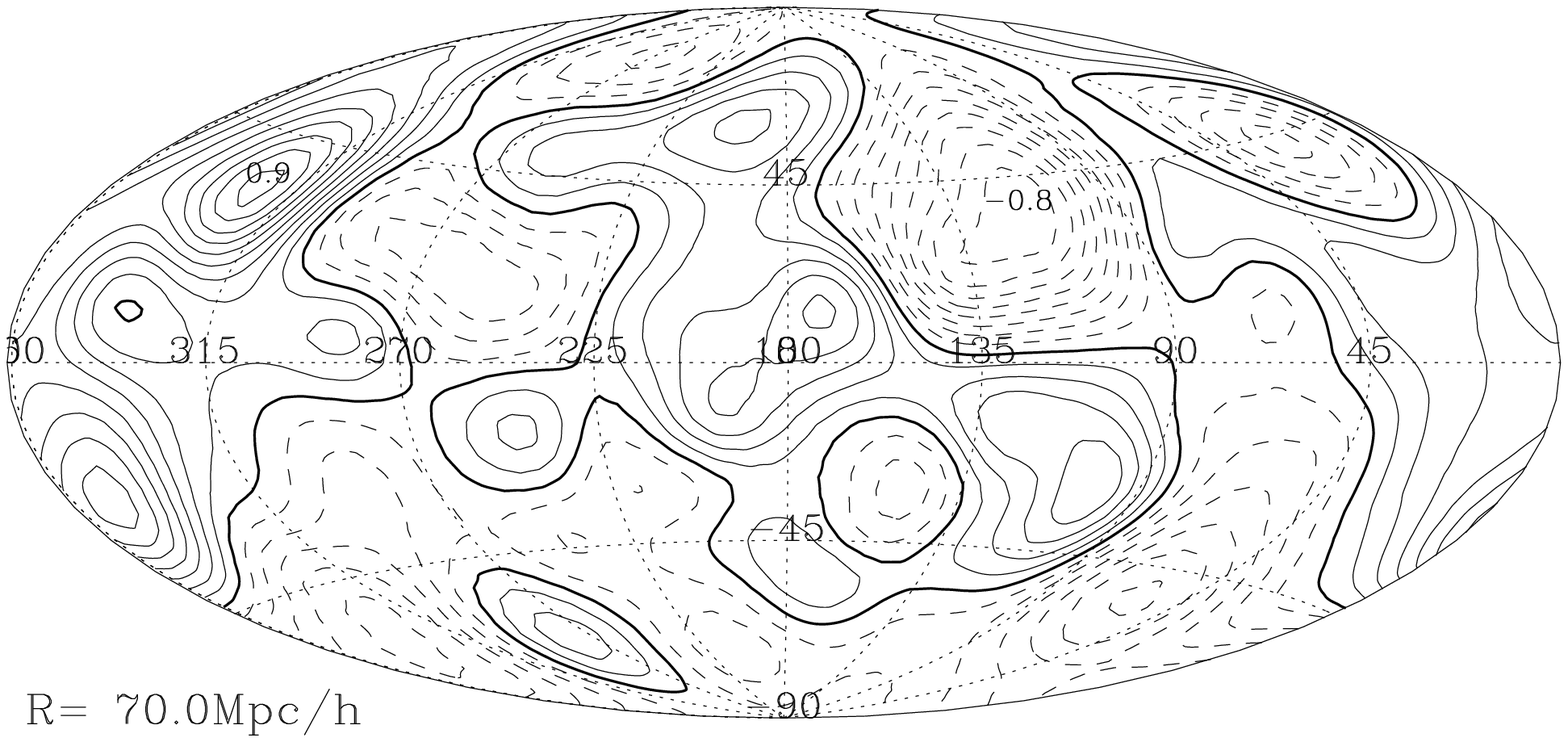}{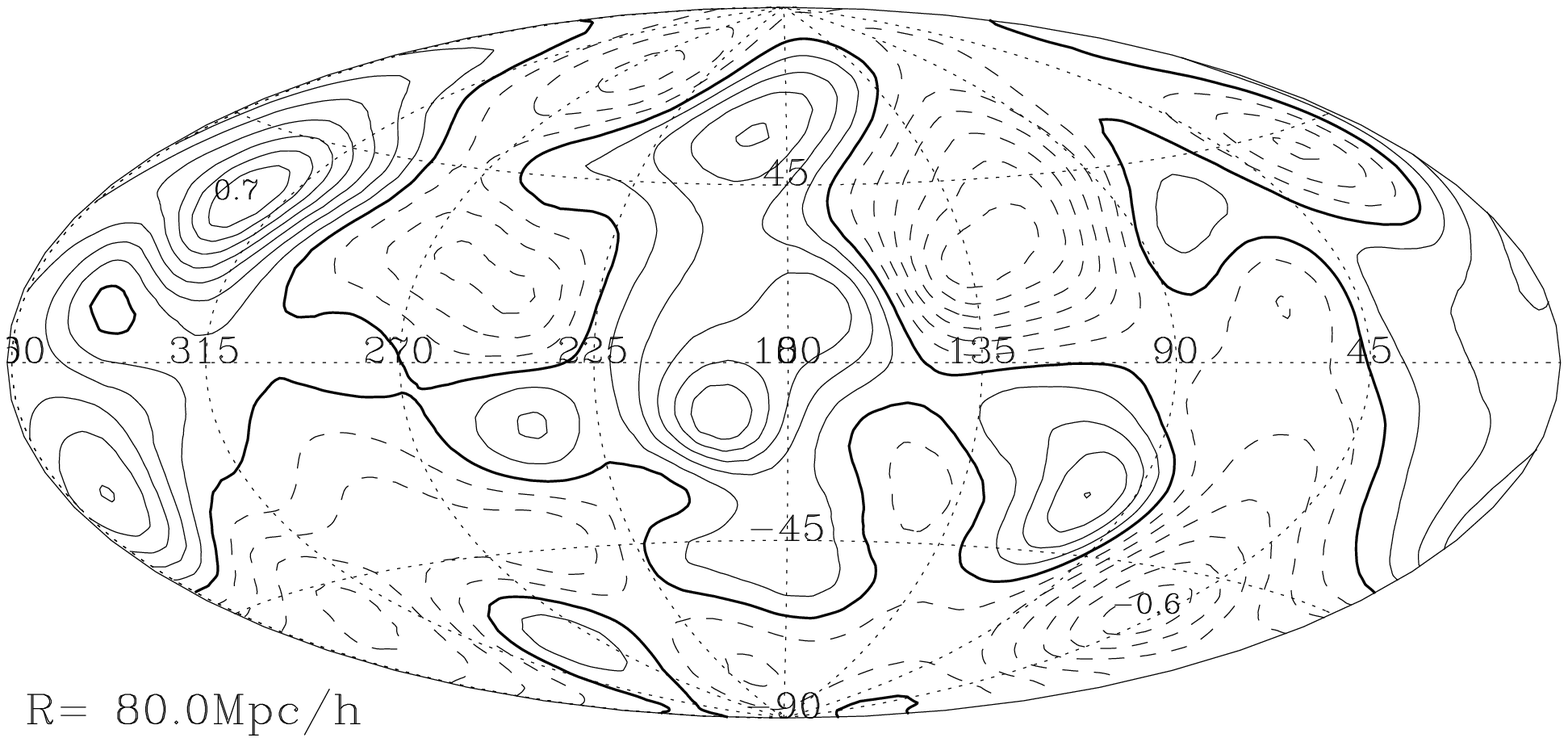}
\caption{The WF reconstructed density maps from the SEcat catalog
evaluated on thin shells at various real space distances, shown in
Galactic Aitoff projections.}
\label{fig:secat_aitoff} 
\end{figure} 

Figure 3 shows the Aitoff projection of the WF SEcat reconstructed
density in Galactic coordinates, evaluated across shells at various
distances.  The structures in these maps match closely those seen in
similar reconstruction from the IRAS 1.2-Jy redshift catalog (Webster
\etal 1997). The $4000\kms$ shell has been discussed earlier, the
other shells will be discussed in detail elsewhere (Zaroubi \etal
2000b)
\begin{figure} 
\plotonea{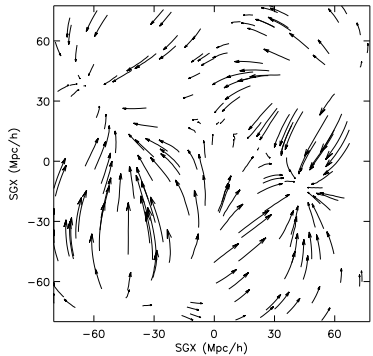} 
\caption{The SEcat reconstructed velocity field, with Gaussian
smoothing of $900\kms$ radius, in the Supergalactic plane is displayed
as flow lines that start at random points, continue tangent to the
local velocity field, and are of length proportional to the magnitude
of the velocity at the starting point.}
\label{fig:streamlines} 
\end{figure} 
The velocity field along the Supergalactic plane is presented in
Fig. 4, showing the existence of two convergence regions which roughly
coincide with the locations of the GA and PP.

\section{Discussion}
In spite of the high level of extinction due to the galactic plane, galaxy
peculiar velocities together with statistical reconstruction
techniques, \eg\, WF, present a very useful tool for mapping the
ZOA. This approach complements the very challenging task of directly
mapping the universe behind the Milky Way.
 
In this contribution we have showed that the WF method could indeed,
within the resolution limit, faithfully reconstruct the nearby
universe including regions masked by the Galactic plane. Several
issues still need to be addressed as the details of the
reconstructed maps can vary from catalog to catalog depending on the
distance indicator, \ie\, TF {\it vs.} \dnsig\, the sampling, noise
properties and systematic differences, \eg\, calibration.

In attempting to combine the ENEAR and SFI data-sets, one needs to
ensure their consistency. Indeed various indications, \eg\,
calibration, zero point, measured bulk flow (da Costa \etal 2000a \&
2000b) support the assumed compatibility of ENEAR and SFI, enabling
their combination to one new catalog of spiral and elliptical
galaxies, SEcat. These issues will be explored in detail in a
forthcoming paper (Zaroubi \etal 2000b).

\acknowledgments I would like to thank M. Bernardi, L.N. da Costa
and my long term collaborator Y. Hoffman for their contribution to
this work and C. Cress for her helpful comments on the manuscrip. The
contribution of the ENEAR team is gratefully acknowledged. The
financial support of the Deutsche Forschungsgemeinschaft (DFG) is
acknowledged.

\end{document}